\documentstyle{mn}                 %
\def\e{{SN~1999E}\/} 
\def\bw{{SN~1998bw}\/} 
\def\cy{{SN~1997cy}\/} 
\def\kms{km s$^{-1}$} 
\def\Ha{H$\alpha$}

\def\m100{mag/100$^d$} 
 
\def\c57{{$^{57}$Co}\/}

\def\ti44{{$^{44}$Ti}\/} 

\input{psfig.tex}

   \title[SN~1999E.]
{SN~1999E. Another piece in the SN--GRB connection puzzle
   \thanks{Based on observations 
 collected at ESO La Silla, Telescopio Nazionale Galileo, OAN-Tonanzintla
 and OAGH-Cananea.}}

\author[Rigon et al.]
{L. Rigon$^{1,2}$, M. Turatto$^{1}$, S. Benetti$^{1}$,
A. Pastorello$^{1,2}$, E. Cappellaro$^{3}$, I. Aretxaga$^{4}$, \and
O. Vega$^{4}$, V. Chavushyan$^{4}$, F. Patat$^{5}$,
I.J. Danziger$^{6}$, M. Salvo$^{7}$
\\   
$^1$ Osservatorio Astronomico di Padova -- INAF, Vicolo dell'Osservatorio 5,
   Padova, I-35122, Italy\\
$^2$ Dipartimento di Astronomia dell'Universita' di Padova, Vicolo
dell'Osservatorio 3, Padova, I-35122, Italy\\
$^3$ Osservatorio Astronomico di Capodimonte -- INAF, Via Moiariello 16,
Napoli, I-80131, Italy\\
$^4$ Istituto Nacional de Astrof\'isica, \'Optica y Electr\'onica,
Aptdo. Postal 51 y 216, Puebla, Mexico\\
$^5$ European Southern Observatory, Karl Schwarzschild Str.2, D-85748,
Garching bei M\"unchen, Germany\\
$^6$ Osservatorio Astronomico di Trieste -- INAF, Via G.B. Tiepolo 11,
Trieste, I-34131, Italy\\ 
$^7$ Australian National University, Mt.Stromlo Observatory, Cotter
Rd, 2611 Weston, ACT, Australia\\
}

\date{Received ................; accepted ................}

\begin{document}

\maketitle

\begin{abstract}

  Detailed optical and near--IR observations of
\e\/ have confirmed early suggestions that this supernova was indeed a
twin of the peculiar type II \cy: it was exceptionally luminous and
had evolved slowly, and the line profiles had narrow peaks and broad wings, 
indicating interaction with the circumstellar material.
Nevertheless, the most intriguing characteristic was that, in analogy
to \cy, it exploded at a position consistent in time and location with
a BATSE event (GRB980910). The "a posteriori" probability that the only
two SNe with such an optical appearance are associated with two
different BATSE GRB
is only 0.2\%. This raises the possibility that some GRB are associated
with H--rich SNe.

\end{abstract} 

%
\section{Introduction} \label{int}

Core collapse supernovae, the final endpoints of massive star evolution,
are usually considered to release about $10^{51}$ ergs
of kinetic energy. In recent years, however, a
new perspective of explosions of a very different energy has emerged. 
On the one hand SN 1997D was successfully modeled with an explosion
energy of only $10^{50}$ ergs (Turatto et al. 1998, Benetti
et al. 2001, Chugai \& Utrobin 2000, Zampieri et al. 2002), whereas
on the other, SNe with explosion energies one order of magnitude
larger than the average have been observed. To this class belong at least
two objects associated (more or less closely) to GRBs, i.e the SNe
1998bw and 1997cy. Iwamoto et al. \shortcite{iwa98bw} have dubbed
supernovae with energy higher than $10^{52}$ ergs as {\em hypernovae},
although the term is not universally adopted.

\bw\/ was an unprecedented object, though with some similarity to SN~1997ef.
Phenomenologically it was first classified as a SNIb \cite{sad98bw}      
and later as a peculiar SNIc \cite{fil98bw}
\cite{patiauc}. This latter classification 
was confirmed by the appearance of the nebular spectrum. 
Expansion velocities
as high as $3 \times 10^4$ \kms\/ and high luminosity \cite{pat98bw}
along with an exceptionally strong radio emission with evidence of
relativistic ejecta, made this object unique \cite{kul98}. 
However, its
most puzzling characteristic was the spatial and temporal association
with the GRB~980425: this SN was discovered inside the $8\arcmin$ error
box of the BeppoSAX WFC centered on the GRB, and the model of the
early light curve indicates that the time of explosion was within
$+0.7/-2$ days of the GRB detection \cite{iwa98bw}.
The exceptional luminosity of \bw\/ and the possible connection with the GRB
(Iwamoto et al. 1998, Woosley et al. 1999) can be
explained either by a hyper--energetic explosion or by a highly
asymmetric emission \cite{hof98bw}.  A comparative
analysis between the data of SN~1997ef and those of \bw\/ has
highlighted significant similarities, although the former has a
slightly smaller explosion energy, $8 \times 10^{51}$ ergs
\cite{iwa97ef}. The recent SN~2002ap seems to lie at an even lower
energy on the hypernovae sequence \cite{mazz02}.

Also there is plausible evidence that a certain number of GRBs at high $z$ 
are due to SNe. Large deviations
from the initial declines observed in the optical afterglows of some GRBs
might be indeed due to SNe.
An interesting case is GRB~011201--SN~2001ke whose afterglow 
optical spectra are consistent with the presence of a core--collapse SN 
shocking a dense CSM \cite{gar01ke}.

Another nearby object, \cy, was possibly associated 
with a GRB \cite{ger97cy}, the uncertainty residing  
in the large error box of the BATSE detection. Indeed, this object showed an
unprecedented photometric and spectroscopic behavior: it was
exceptionally luminous, possibly the most luminous SN ever detected, and slowly
declining, while the spectra showed signs of ejecta-CSM interaction at all
epochs \cite{tur97cy}. The entire light curve has
been reproduced by a model that requires the interaction of very energetic
ejecta ($3 \times 10^{52}$ ergs) of a 25 M$_\odot$ massive star with the
CSM (with some contribution from radioactive decay).

In this paper, we report on the properties of \e\/ which showed many
of the properties of \cy\/ (Cappellaro et al. 1999,
Filippenko 1999b, Filippenko 2000). Indeed,
in addition to a remarkable similarity in the optical domain, a loose
but intriguing spatial and temporal coincidence with a GRB was
noted by Thorsett \& Hogg \shortcite{thor}: in the span between this SN's
discovery and the previous known observation of the parent galaxy (1998
Jul 29.046 U.T.), GRB980910 was detected at a position spatially
consistent with that of \e\/ (separation of $4.8$\degr with a
positional uncertainty $\sigma=6.8$\degr~for the GRB).

   \begin{figure}
   \psfig{figure=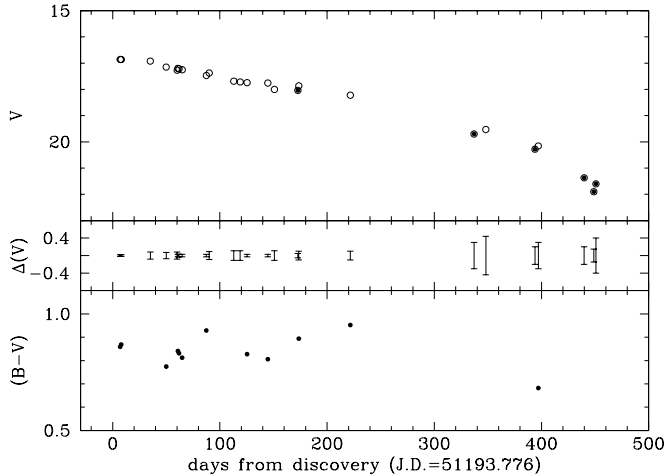,angle=-90,width=9cm} 
   \caption{Top: V
   light curve of \e. Open symbols represent measurements performed
   with the PSF fitting technique, filled symbols those performed with
   the background subtraction method. Middle:
   estimates of the errors of the photometry performed via artificial
   star experiments. Bottom: (B-V) color curve of \e. }
   \label{sn99e_ph} \end{figure}
    
\section{\e\/} \label{99e}

\e\/ was discovered by Antezana on Jan. 15, 1999 \cite{99eiauc}
0\farcs9 West and 10\farcs0 South of the nucleus of an
anonymous galaxy at R.A. = 13h17m16\fs37, Decl. = $-18$\degr33\arcmin13\farcs4
(equinox 2000.0). The galaxy redshift is $z_{LSR}=0.0261$ ($\mu=35.40$ adopting
$H_O=65$ \kms Mpc$^{-1}$) as measured on the narrow 
peaks of the Balmer emission lines of the SN.

\e\/ has been classified as a type IIn due to the presence of a strong and 
narrow H$\alpha$ emission line (Filippenko et al. 1999, Jha et
al. 1999).

  \begin{figure}
   \psfig{figure=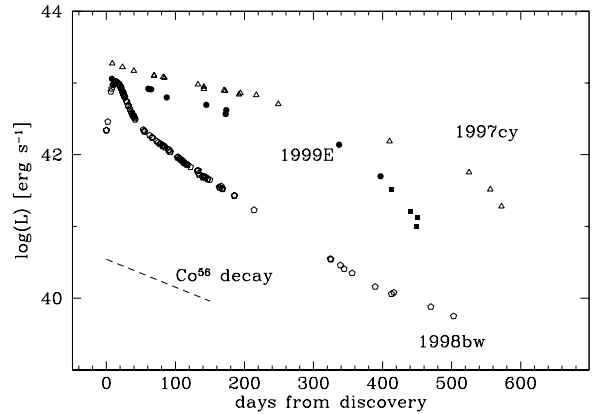,angle=-90,width=9cm}
   \caption{UBVRI bolometric light curve of \e\/ 
   compared to those of the hypernovae 1998bw (Patat et al. 2001)
   and 1997cy (Turatto et al. 2000). For each SN the luminosities are
   plotted versus the days from discovery. The last estimates have
   been scaled from the V magnitudes only. 
   The dashed line corresponds to the $^{56}$Co to $^{56}$Fe decay rate,
   expected for full $\gamma$--ray trapping. 
   }
    \label{sn99e_bol1}
    \end{figure}

\subsection{The light curves} \label{lc}
Despite the unfavorable location of the SN upon a spiral arm, \e\/ has
been extensively observed in the optical (UBVRI) bands for over 450
days at different observatories (ESO La Silla: Dutch, Danish+DFOSC,
3.6+EFOSC2; La Palma: TNG; Mexico: OAN-Tonanzintla and
OAGH-Cananea). The SN magnitudes were measured through PSF fitting
technique. In order to check for the amount of background
contamination at late phases we derived the SN
magnitudes on some frames with the template subtraction technique (see
Fig. \ref{sn99e_ph}). The PSF technique seems to be adequate at least
up to 200d after discovery. However,  in order to obtain precise
determination of the magnitudes at the latest phases, we subtracted 
from the SN
images the reference frames of the parent galaxy taken 2.5 yr after
the explosion, when the SN had faded. A few JHK' observations are also
available (ESO La Silla: NTT+SOFI). The V light curve and the (B-V)
color curve are shown in Fig.~\ref{sn99e_ph}.  The object closely
resembles the behavior of type IIn SNe like SN~1988Z \cite{tur88z}, 
with a slow luminosity decline and an
almost constant color.

The optical photometric data have been used to get the UBVRI
"bolometric" light curve, shown in Fig.~\ref{sn99e_bol1}.  Infrared
data have been excluded because they were available only in three
nights.  We have checked, however, that in all of these nights their
contribution to the total (optical and infrared) luminosity was less
than 10\%. In the same figure we also show the
bolometric light curves of SNe 1997cy and 1998bw derived in the same
way, for comparison.

Because of the presence of significant interstellar absorption lines
in the spectra, it is reasonable to expect \e\/ to be significantly
reddened.  The EW of the NaID absorption lines originating in the
parent galaxy, as measured in the spectra of higher resolution and
signal-to-noise ratio, is about twice as strong as the Galactic component.
Assuming that the gas to dust ratio inside the parent galaxy is the
same as that in the Galaxy, we have adopted for the \e\/ a total
absorption $ A_{B,total}=1.14$ mag, as the sum of $A_{B,MW}=0.38$
\cite{schl98} and $A_{B,host}=2 \times
A_{B,MW}$. With such an absorption value  the absolute magnitude
becomes M$_V <-19.5$, even ignoring the uncertain early magnitudes reported
in the IAU Circulars.

Whatever the absorption suffered by \e\/ is, the bolometric
light curve shape differs from those of other type II
SNe: it does not show (at least up to about 400 days after the 
discovery) the radioactive tail
due to $^{56}$Co which characterizes normal SNII 
after 100 days from explosion. By comparison with SN~1987A,
the bolometric light curve provides  
an upper limit to the $^{56}$Co mass of about 1M$_\odot$. 
However, by analogy with other SNIIn, e.g. SNe 1988Z (Turatto et al. 1993, 
Aretxaga et al. 2001) and
1995G \cite{apas}, it is natural to attribute 
the slow decline of the light curve to interaction with a dense CSM.
In particular, the light curve of \e\/  appears
very similar to that of \cy. Like the latter, \e\/ is a very energetic
supernova, with an explosion luminosity a few days after discovery exceeding
$10^{43}$ erg s$^{-1}$. The luminosity of \cy\/ was even higher
($1.86\times10^{43}$ erg s$^{-1}$), although it should be noted that for
both SNe the epochs of maxima were not well known.

\subsection{The spectrum}{\label{sect.sp}
Spectra of \e\/ at different epochs have been obtained at ESO La Silla
(1.5m+B\&C, 3.6m+EF2) and La Palma (WHT+ISIS) (see
Fig.~\ref{evol_spec}). The spectra show very broad features indicating
very high expansion velocities, and the similarity with \cy\/ is striking.
This is illustrated in Fig.~\ref{sp:sn99e_97cy} where the spectra
of the two SNe at two different epochs are compared.  Even though
evolution is slow, remarkably enough, the best correspondence between
the two SNe is obtained by computing the phase from the respective GRB.

Figure~\ref{sn99e_ha} shows the evolution of the \Ha\/ profile of \e, 
on the basis of our spectra with higher resolution.
When the resolution is better than 4 \AA\/ (FWHM),
a narrow P--Cygni profile can be detected above 
the broad emission. An underlying HII region, 
which could not be
properly subtracted at the last two epochs (note the residual emissions
of [NII], [SII] and \Ha\/ on July 20), contributes to
the unresolved emission.
However, the intensity of the narrow emission line 
decreases with time, indicating the presence of 
a component associated with the SN.

As in other SNIIn, e.g. SNe 1988Z \cite{tur88z}, a
good fit to the \Ha\/ line profiles requires at least 3 components. At
the first epoch (Jan 23, 1999) the 3 components have gaussian profiles
with FWHM=8600, 1900 and $200$ \kms, respectively. The presence of
P--Cygni profiles indicates the existence of a dense, slow--moving
wind around the exploding star, like in other type IIn SNe, e.g. SNe
1997ab (v=180 \kms, Salamanca et al. 1998) and 1995G (v=900
\kms, Pastorello et al. 2002).  The positions of the absorption
minima in \e\/ indicate an expansion velocity of this shell of about
200 \kms.

Line identifications in the spectra of SNe with high expansion
velocities are difficult. An attempt was made for \e\/ by Filippenko
\shortcite{fil} in the optical spectra of the first months. In
Fig.~\ref{sn99e_id} we identify the major lines in the optical--NIR
spectrum of \e\/ on March 1999. Balmer and Paschen lines of hydrogen
are clearly detected: the Paschen lines show the same multicomponent
structure as the Balmer ones. Also the Ca~II IR triplet and H \& K are
clearly visible. O~I 8446 \AA\/ is identified on the blue wing of the
Ca~II IR triplet and possibly O~I 7774 \AA\/ is detected as well. 
The Pa 6 line is blended with He~10830 \AA.

  \begin{figure*}
    \psfig{figure=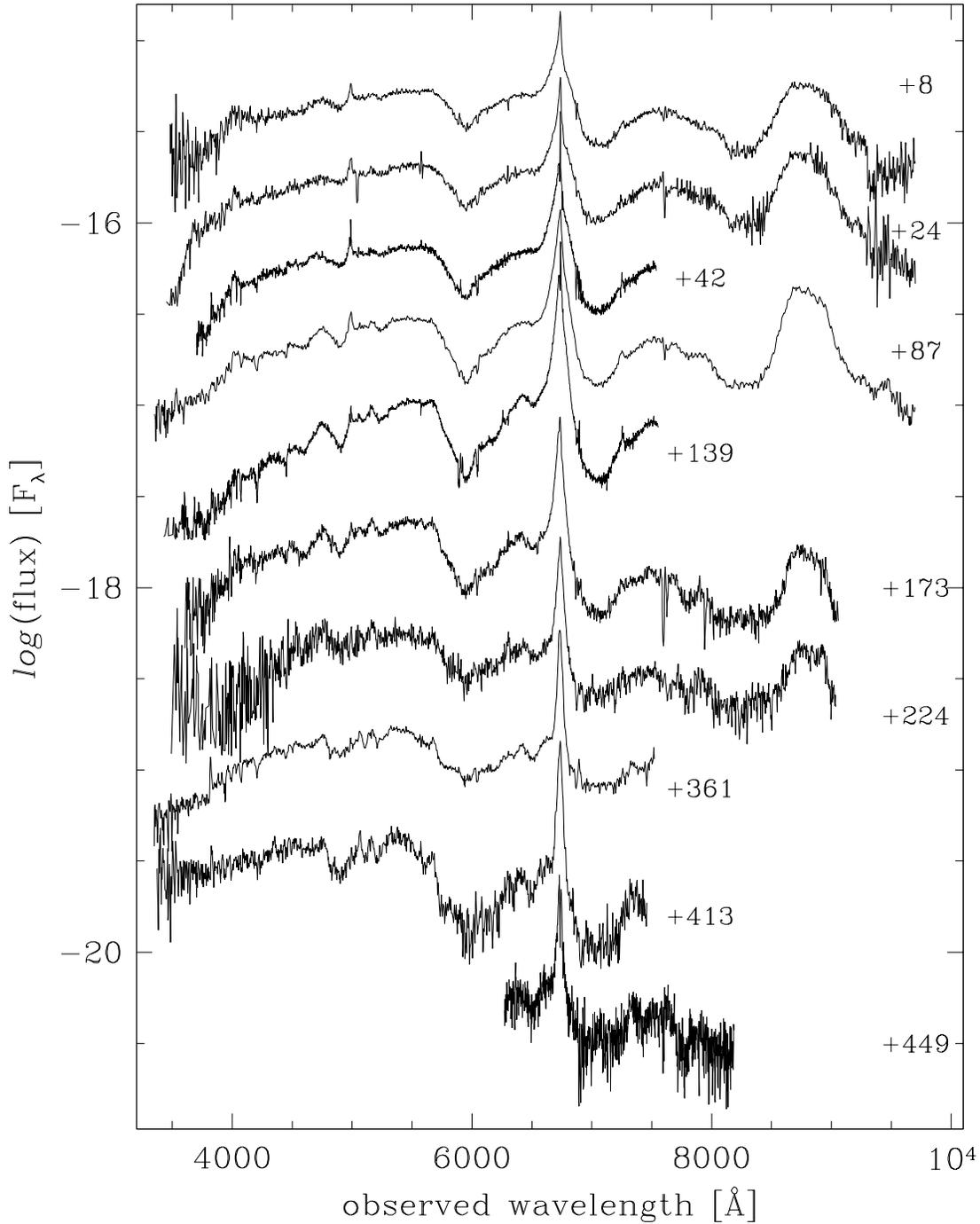,angle=0,width=15cm} 
  \caption{ Spectral evolution of SN~1999E. The ordinate refers to the first
  spectrum, and the others have been arbitrarily shifted. The phase (days) is
  from the discovery date (Jan. 15, 1999). The spectra shown have been obtained
  with the following instruments:
  phase +8d, Danish 1.5m telescope at La Silla + DFOSC, resolution 12\AA;
  +24d, ESO 1.5 +B\&C, res 15\AA; +42d, William Hershel telescope at La Palma +ISIS, 
  res 3.1\AA;
  +87d, ESO 3.6m + EFOSC2, res 17\AA; +139d, ISIS, res 3.1\AA; +173d, DFOSC, res 12\AA;
 +224d, DFOSC, res 12\AA; +361d, EFOSC2, res 18\AA; +413d, EFOSC2, res 18\AA; +449d,
 EFOSC2, res 6.5A.  
  }  \label{evol_spec} 
  \end{figure*}

\section{SN1999E and GRB980910} \label{sect_grb}

The possibility that some Gamma Ray Bursts (GRBs) are associated 
with SN explosions has been suggested in recent years
with the discovery of a handful of noteworthy
temporal and spatial coincidences
between GRBs and SNe, e.g. GRB980425/SN~1998bw \cite{galama},
GRB970514/SN~1997cy \cite{ger97cy}, GRB971115/SN~1997ef 
\cite{wang98} and GRB011121/SN~2001ke (Garnavich et al. 2002, Price et al. 2002,
Bloom et al. 2002).

\e\/ can be added to this list.
Thorsett \& Hogg \shortcite{thor} noted that GRB980910 (BATSE trigger n.7077)
was detected only 4.8\degr (corresponding to 0.73$\sigma$) from the SN
location in the  time-interval elapsed
between the discovery and the last pre-SN image of the host
galaxy. Telemetry gaps prevented the inclusion of this burst in the flux and
duration tables (Meegan C., private communication 2002). In that 172 day time
interval there were 98 GRBs recorded in the BATSE GRB catalog 
\footnote{http://www.batse.msfc.nasa.gov/batse/grb/catalog/}, 
including both overwrite and
overwritten events \cite{batse}. 
Among these, GRB980910 is the one closest to the \e\/
location. Another burst, GRB 980920,  with a much larger error box (1
$\sigma=9^o.98$), lies at about $1.2\sigma$.

The probability of a chance association of \e\/ with one of the 98
GRBs at a 0.73 $\sigma_i$ distance is 9.6\%, where $\sigma_i$ is the
corresponding error circle of each GRB (see Wang \& Wheeler 1998,
Germany et al. 2000). The same computation was performed for
\cy. During the 126 days elapsed between the discovery of \cy\/ and
the previous image of the host galaxy which does not show trace of the
SN, 125 GRBs were detected. GRB970514, at 0.23$\sigma$ distance of the
SN location (1$\sigma=3^o.8$), was the closest. The probability
of a chance association at a 0.23$\sigma_i$ distance from \cy\/ is
only 1.8\%.

 \begin{figure}
   \psfig{figure=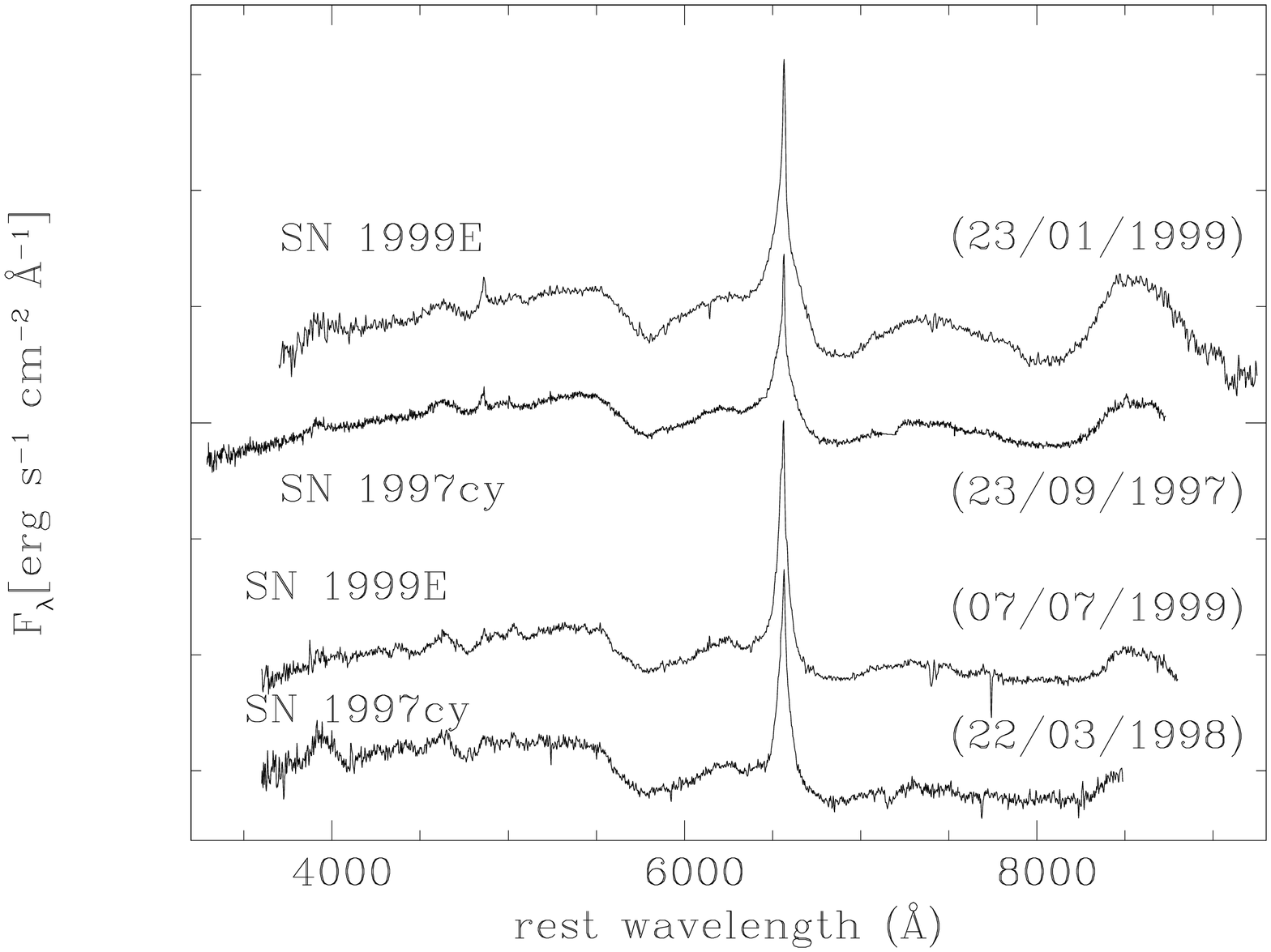,angle=-90,width=9cm}
 \caption{Comparison of the spectra of \e\/ and \cy,
 corrected to the parent galaxy rest frames. 
 The two spectra at the top have been obtained about 135 days past the
 GRB to which the SNe are possibly associated (+8d past discovery for \e); 
 those at the bottom
 about 300 days (+173 past discovery for \e). The ordinate scale refers
 to the top spectrum of \e, other spectra have been arbitrarily 
 downshifted. The bottom two spectra have also been multiplied by a
 factor 2 for an easier comparison.}
    \label{sp:sn99e_97cy}
    \end{figure}   

From the probability computed above it appears that the association 
GRB980910/SN~1999E is less stringent than for \cy.
However, in the previous Sections we have noted 
the striking similarities in luminosity and spectral evolution
between \e\/ and \cy. To our knowledge, no other SNe observed so far have 
shared the same properties as these two objects. Therefore, one might argue
in favor of a direct relation between those
SNe with such optical features and $\gamma$--ray
emission.

If we combine the probability of association of both SNe with their
respective GRBs, we end up with only a 0.2\% probability of chance
association.  Although we are aware of the caveats on "a posteriori"
statistics, formally there is a 99.8\% probability of a causal link
between SNe like \cy\/ and \e\/, and some kinds of GRBs.

   \begin{figure}
   \psfig{figure=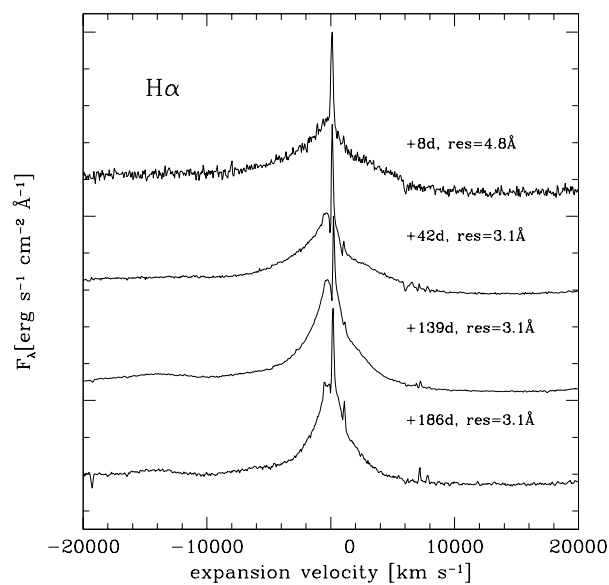,angle=0,width=9cm}
   \caption{Time evolution of the \Ha\/ profile from the spectra of higher 
   resolution. The line fluxes have been normalized to the peak intensities.
   Broad wings and narrow unresolved
   emissions and absorptions are visible.}
   \label{sn99e_ha}
   \end{figure}   

   \begin{figure}
   \psfig{figure=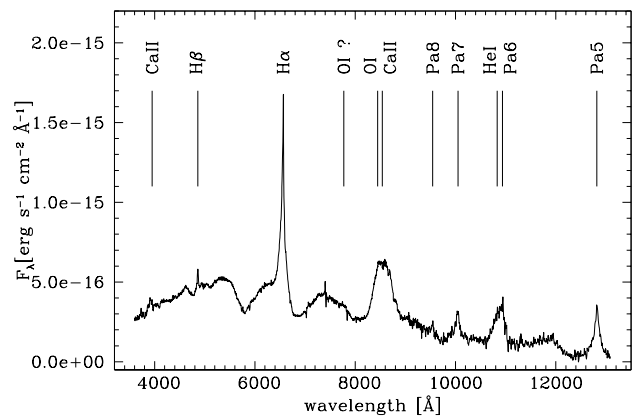,angle=-90,width=9cm}
   \caption{Line identifications in the optical--IR spectrum of \e\/ 
   (taken on March 11, 1999 with 1.5m+B\&C, phase +55d, and March 7, 
   1999 with NTT+Sofi, phase +51d)
   shifted to the rest frame of the host galaxy.}
   \label{sn99e_id}
   \end{figure}

\section{Conclusions} \label{conc}
\e\/ was an intrinsically luminous (M$_V <-19.5$) type II SN. The
light curve, flatter than the $^{56}$Co decline during the first 400
days, requires an additional source of luminous energy, most likely
resulting from the interaction with a dense CSM. The CSM signature can
be read in the narrow ($<200$ \kms) P-Cygni profile detected in our
highest resolution spectra. The color evolution of the SN also differs
from that of normal SNII, which soon after the maximum undergoes a phase
of cooling (and reddening) due to expansion (cfr. Patat et al. 1994). 
Indeed the spectra also
show prominent lines without the characteristic P--Cygni profile and
with broad wings (see Fig. \ref{evol_spec}), as in other type IIn SNe.

The most intriguing features of \e\/ are its possible association with
GRB~980910 and its striking similarity to \cy, also associated with a
GRB.  In Sect.~\ref{sect_grb} we have argued that the probability of a
chance association of these two SNe with GRBs is low. We have also
found that a remarkable correspondence between the spectral evolutions
of the two SNe can be found if the phases are computed from the epoch
of the GRB (cfr. Fig.~\ref{sp:sn99e_97cy}).

The epoch of explosion of \e\/ is not known and poorly constrained
\cite{99eiauc} as is that of \cy. It is natural to
compare the light curves of these two SNe, taking the GRB trigger times
as epoch of the explosions. The result of this comparison
is shown in Fig.~\ref{sn99e_bol2}: up to day 460 after the GRBs 
there is substantially no distinction between both energetics and
time evolution of \cy\/ and \e!  
After that epoch, \e\/ starts a fast decline, while for
\cy\/ this occurs 100 days later.

The observations of \e\/ seem, therefore, to support the conclusions drawn
from \cy\/ that these SNe are very energetic (E$_{tot}\sim 10^{52}$ ergs)
and may have massive progenitors \cite{tur97cy}. The
superposition of both SNe on spiral arms of their respective host galaxies is
consistent with this conclusion.
The forbidden lines typical of the late time spectra of core collapse
SNe are not seen in the spectra of \e\/, even  in those obtained more 
than 1 year  after the discovery, probably 
overwhelmed by emission due to the ejecta--CSM interaction.
The simple detection of the OI lines, obtained thanks to the high 
signal--to--noise 
reached in the observations of \e\/ (cfr. Sect.
\ref{sect.sp} and Filippenko \shortcite{fil}), 
cannot tell us much about oxygen produced in the explosion, nor whether
any has been drawn into black hole formation.  In fact, the oxygen
producing this line may well be that which existed in the envelope of
the progenitor star.  The interaction of the SN ejecta with a dense
CSM powers the light curves up to 460 (\e) and 580 (\cy) days after the burst.
The difference in the subsequent drop of luminosity is
possibly due to different density distribution of the CSM.

The close similarity of the two events, both in terms of optical 
SN and coincidence of 
$\gamma-$ ray bursts, supports the hypothesis that this kind
of energetic SNe may be associated with some specific type of GRB.
Unfortunately, little is known about GRB980910
(Sect.\ref{sect_grb}). On the contrary we do know that GRB970514,
associated to \cy, was of short duration ($\sim 0.2$s) and the total
energy emitted was $\sim 10^4$ times less than those of other bursts
with measured redshift \cite{ger97cy}. The collapsar model explains
the properties of the GRBs with long duration through the collapse of
massive stellar cores into black holes \cite{woos} but does not
account for bursts as short as GRB970514. This model requires the
core to be massive enough to form a black hole and with a large angular
momentum that forms a disk. It also requires the H and He envelopes
of the star to have been stripped long before the explosion.  This is not the
case of \e\/ and \cy, for which there is evidence of a dense, low-velocity 
shell of H--rich gas
associated with the SN.\\ 
The supranova model \cite{vietri}, based on
the delayed implosion of a rotating neutron star into a black hole,
associates GRB with SNe from massive stars. In this scenario the SN
explosion take place months or years before the GRB because the SN
remain optically thick to $\gamma-$rays for several months. Because of
the negative observations available for the two SNe about two months
before the GRB, only a small temporal window for the explosions is
available. This might be too tight a constraint.

Despite there not being a straightforward explanation for the 
explosions of \e\/ and \cy\/ within the current theoretical models,
we find the possibility of an association between these 
H--rich, high--energy, CSM--interacting SNe with some GRB attractive.

   \begin{figure}
   \psfig{figure=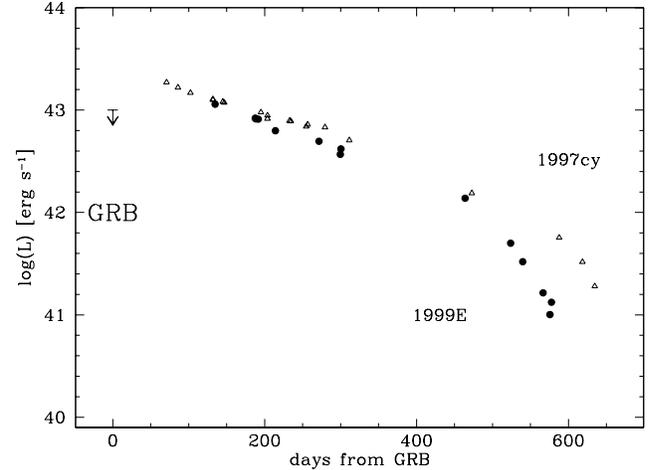,angle=-90,width=9cm}
   \caption{Comparison of the UBVRI bolometric light curve of \e\/ 
   with that of \cy\/ when the phase is computed 
   from the respective GRB.}
   \label{sn99e_bol2}
   \end{figure}   

\bigskip

\noindent
{\bf ACKNOWLEDGMENTS}
We are indebted to A. Pizzella for having provided the reference frame
of the parent galaxy used for the image subtraction, and to Janet Clench
for correcting the manuscript.
 We acknowledge support from the Italian Ministry for University and 
 Research (MIUR) through grant Cofin MM02905817. We are grateful to
 Charles Meegan for having looked in detail into the BATSE data of 
 GRB980910. This work has been partly supported by CONACyT grant E-3218.

\end{document}